\documentclass[a4paper,twoside,reqno,11pt,tbtags]{article}

\usepackage{amssymb,amsmath,amsthm,mathrsfs}
\usepackage{graphicx}
\usepackage{times}
\usepackage[round]{natbib}

\leftmargini=\baselineskip

\newtheoremstyle{localthm}
	{5pt} 
	{5pt} 
	{\sl} 
	{} 
	{\bf} 
	{{\rm.}} 
	{.7em} 
	{} 

\theoremstyle{localthm}

\newtheoremstyle{localrem}
	{5pt} 
	{5pt} 
	{\rm} 
	{} 
	{\bf} 
	{{\rm.}} 
	{.7em} 
	{} 

\theoremstyle{localrem}

\def\bs{\boldsymbol}
\def\bDelta{\bs{\Delta}}
\def\bPi{\bs{\Pi}}
\def\bSigma{\bs{\Sigma}}
\def\A{\bs{A}}
\def\C{\bs{C}}
\def\u{\bs{u}}
\def\U{\bs{U}}
\def\v{\bs{v}}
\def\V{\bs{V}}
\def\W{\bs{W}}
\def\w{\bs{w}}
\def\x{\bs{x}}
\def\y{\bs{y}}
\def\z{\bs{z}}
\def\R{\mathbb{R}}

\renewcommand{\Pr}{\operatorname{\mathbb{P}}}
\newcommand{\diag}{\operatorname{\mathrm{diag}}}

\begin{document}

\title{Refining Invariant Coordinate Selection via Local Projection Pursuit}
\author{Lutz D{\"u}mbgen, Katrin Gysel and Fabrice Perler\\
	University of Bern}
\date{\today}

\maketitle

\begin{abstract}
Invariant coordinate selection (ICS), introduced by \cite{Tyler_etal_2009}, is a powerful tool to find potentially interesting projections of multivariate data. In some cases, some of the projections proposed by ICS come close to really interesting ones, but little deviations can result in a blurred view which does not reveal the feature (e.g.\ a clustering) which would otherwise be clearly visible. To remedy this problem, we propose an automated and localized version of projection pursuit (PP), cf.\ \cite{Huber_1985}. Precisely, our local search is based on gradient descent applied to estimated differential entropy as a function of the projection matrix.
\end{abstract}

\noindent
$^*$Work supported by Swiss National Science Foundation.\\

\addtolength{\baselineskip}{0.1\baselineskip}

\section{Projection pursuit}
\label{sec:projection.pursuit}

Suppose our data consists of vectors $\x_1, \x_2, \ldots, \x_n \in \R^p$, and we view these as independent copies from a random vector $\bs{X}$ with unknown distribution $P$. If $p \in \{1,2,3\}$, there are various ways to display the data graphically and find interesting structures, e.g.\ two or more separated clusters or manifolds close to which the data are concentrated. If we use a particular method to visualize data sets in dimension $d \in \{1,2,3\}$ but $p > d$, we want to find a $d$-dimensional linear projection of the data which exhibits interesting structure. This task has been coined ``projection pursuit'' (PP) by \cite{Friedman_Tukey_1974}. We also refer to the excellent discussion papers of \cite{Huber_1985} and \cite{Jones_Sibson_1987} for different aspects and variants of this paradigm.

More formally, if the distribution $P$ has been standardized already in some way, our goal is to determine a matrix $\A \in \R^{p\times d}$ with orthonormal columns such that the distribution $P_{\A}$ of $\A^\top\bs{X}$ is ``interesting''. In what follows, such a matrix $\A \in \R^{p\times d}$, that is, $\A^\top \A = \bs{I}_d$, is called a ``($d$-dimensional) projection matrix'', and the distribution $P_{\A}$ is called a ``projection of $P$ (via $\A$)''.

An obvious question is how to measure whether a distribution $Q$ on $\R^d$ is ``interesting''. To answer this, let us summarize some of the considerations of \cite{Huber_1985}. Suppose that $Q$ has density $g$ with respect to $d$-dimensional Lebesgue measure. Shannon's (differential) entropy of $Q$ is defined as
\[
	H(Q) \ := \ - \int \log(g(\y)) g(\y) \, d\y .
\]
It is well-known that among all distributions $Q$ with given mean vector $\bs{\mu} \in \R^d$ and nonsingular covariance matrix $\bSigma \in \R^{d\times d}$, the Gaussian distribution $\mathcal{N}_d(\bs{\mu}, \bSigma)$ is the unique maximizer of $H(Q)$. Coming back to the distribution $P$, that its projection $Q = P_{\A}$ is non-interesting if it is Gaussian is also supported by the so-called Diaconis--Freedman effect, cf.\ \cite{Diaconis_Freedman_1984}. Under mild assumptions on $P$, most projections $P_{\A}$ look Gaussian. Precisely, if $\A$ is uniformly distributed on the manifold of all $d$-dimensional projection matrices, then for fixed $d$,
\[
	P_{\A} \ \to_{w,\Pr} \ \mathcal{N}_d(\bs{0},\bs{I}_d)
	\quad\text{as} \ p \to \infty, \ p^{-1} \|\bs{X}\|^2 \to_{\Pr} 1, \ p^{-1} \bs{X}^\top \tilde{\bs{X}} \to_{\Pr} 0 , 
\]
where $\tilde{\bs{X}}$ is an independent copy of $\bs{X}$; see also \cite{Duembgen_Zerial_2013}.

In view of these considerations, a possible strategy is to find a projection matrix $\A$ such that $\hat{H}(\A^\top \x_1,\ldots,\A^\top \x_n)$ is minimal, where $\hat{H}(\y_1,\ldots,\y_n)$ is an estimator of $H(Q)$, based on observations $\y_1,\ldots,\y_n \in \R^d$.

In the present paper, we focus on the entropy $H(Q)$ and the estimated entropy $\hat{H}(\y_1,\ldots,\y_n)$. As discussed by \cite{Huber_1985}, \cite{Jones_Sibson_1987} and many other authors, there are numerous proposals of ``PP indices'' measuring how interesting a distribution $Q$ or the empirical distribution of $\y_1,\ldots,\y_n$ is. As explained at the end of Section~\ref{sec:local.optimization}, our local search method can be easily adapted to different PP indices.

\section{Invariant coordinate selection as a starting point}
\label{sec:ICS}

Invariant coordinate selection (ICS), introduced as a generalization of independent component analysis by \cite{Tyler_etal_2009}, may be described as a two-step procedure. In a first step, the centered raw observations $\x_1^{\rm raw},\ldots,\x_n^{\rm raw}$ are standardized by means of some scatter estimator $\hat{\bSigma}_0 = \hat{\bSigma}_0(\x_1^{\rm raw},\ldots,\x_n^{\rm raw}) \in \R^{p\times p}_{{\rm sym},+}$, where $\R^{p\times p}_{{\rm sym},+}$ stands for the set of symmetric, positive definite matrices in $\R^{p\times p}$. Having determined $\hat{\bSigma}_0 = \bs{B}_0^{} \bs{B}_0^\top$, we replace the raw observations $\x_i^{\rm raw}$ with the standardized observations $\x_i := \bs{B}_0^{-1} \x_i^{\rm raw}$. Strictly speaking, these standardized observations $\x_i$ are no longer stochastically independent, but this is not essential for the subsequent considerations.

To the preprocessed observations $\x_i$ we apply a different estimator of scatter to obtain $\hat{\bSigma} = \hat{\bSigma}(\x_1,\ldots,\x_n) \in \R^{p\times p}_{{\rm sym},+}$. Now we compute a spectral decomposition $\hat{\bSigma} = \sum_{i=1}^p \hat{\lambda}_i \hat{\u}_i^{} \hat{\u}_i^\top$ with eigenvalues $\hat{\lambda}_1 \ge \cdots \ge \hat{\lambda}_p > 0$ and an orthonormal basis $\hat{\u}_1, \ldots, \hat{\u}_p$ of $\R^p$. Then the resulting invariant coordinates correspond to the mappings
\[
	\R^q \ni \x^{\rm raw} \ \mapsto \ \hat{\u}_k^\top \bs{B}_0^{-1} \x_{}^{\rm raw} \in \R ,
	\quad \ 1 \le k \le p .
\]
The results of \cite{Tyler_etal_2009} suggest to look at the $d+1$ particular projection matrices
\begin{align*}
	\A_0 \ &:= \ \bigl[ \hat{\u}_1 \, \ldots \, \hat{\u}_d \bigr] , \\
	\A_1 \ &:= \ \bigl[ \hat{\u}_1 \, \ldots \, \hat{\u}_{d-1} \, \hat{\u}_p \bigr] , \\
	\vdots & \quad \vdots \\
	\A_d \ &:= \ \bigl[ \hat{\u}_{p-d+1} \, \ldots \, \hat{\u}_p \bigr] ,
\end{align*}
that is, the columns of $\A_k$ are the vectors $\hat{\u}_i$ with $1 \le i \le d-k$ or $p-k < i \le p$. One could also consider all $\binom{p}{d}$ matrices
\[
	\A \ = \ [\hat{\u}_{i(1)} \, \cdots \, \hat{\u}_{i(d)}]
	\quad\text{with}\quad
	1 \le i(1) < \cdots < i(d) \le p .
\]

\section{Estimation of entropy}
\label{sec:entropy}

For given observations $\y_1,\ldots,\y_n \in \R^d$ with unknown distribution $Q$ and density $g$, a standard estimator of $g$ would be a kernel density estimator with standard Gaussian kernel,
\[
	\hat{g}_h(\y) \ := \ n^{-1} \sum_{j=1}^n \phi_h(\y - \y_j) ,
\]
where $\phi_h(\y) := h^{-d} \phi(h^{-1} \y)$ with $\phi(\y) := (2\pi)^{-d/2} \exp( - \|\y\|^2/2)$, and $h = h(n) > 0$ is some bandwidth to be specified later. Then a possible estimator of $H(Q)$ is given by
\[
	\hat{H}(\y_1,\ldots,\y_n) \
		:= \ - n^{-1} \sum_{i=1}^n \log \hat{g}_h(\y_i) .
\]
Note that $\hat{H}(\cdot)$ is continuously differentiable with
\begin{align*}
	\hat{H}(\y_1 + \bs{\delta}_1, \ldots, \y_n + \bs{\delta}_n) \
	= \ &\hat{H}(\y_1,\ldots,\y_n) \\
		&+ \ n^{-1} h^{-2} \sum_{i=1}^n
		\frac{\sum_{j=1}^n (\bs{\delta}_i - \bs{\delta}_j)^\top(\y_i - \y_j) \phi_h(\y_i - \y_j)}
			{\sum_{j=1}^n \phi_h(\y_i - \y_j)} \\
		&+ \ O \bigl( \|\bs{\delta}_1\|^2 + \cdots + \|\bs{\delta}_n\|^2 \bigr)
\end{align*}
as $\bs{\delta}_1, \ldots, \bs{\delta}_n \to \bs{0}$, because $\phi_h(\y + \bs{\delta}) = - h^{-2} \bs{\delta}^\top\y \phi_h(\y) + O(\|\bs{\delta}\|^2)$ as $\bs{\delta} \to \bs{0}$. This expansion will be useful for local optimization.

The smoothing parameter $h$ has an impact, of course. Suppose that the underlying distribution $Q$ is the standard Gaussian $\mathcal{N}_d(\bs{0},\bs{I}_d)$. Then the expected value of $\hat{g}(\y)$ equals $\phi_{(1 + h^2)^{1/2}}(y)$, the density of the convolution of $Q$ and $\mathcal{N}_d(\bs{0}, h^2 \bs{I}_d)$. Hence, $\hat{H}(\y_1,\ldots,\y_d)$ may be viewed as an estimator of
\begin{equation}
\label{eq:standard.value.H}
	- \int \log \phi_{(1 + h^2)^{1/2}}(\y) \phi(\y) \, d\y
	\ = \ (d/2) \bigl( (1 + h^2)^{-1} + \log(1 + h^2) + \log(2\pi) \bigr) .
\end{equation}

\section{Local optimization}
\label{sec:local.optimization}

For our purposes it is convenient to over-parametrize the search problem by writing
\begin{equation}
\label{eq:A.via.U}
	\A \ = \ \U \bPi
\end{equation}
with an orthogonal matrix $\U \in \R^{p\times p}$ and the standard projection matrix
\[
	\bPi \ := \ \begin{bmatrix} \bs{I}_d \\ \bs{0} \end{bmatrix} \ \in \ \R^{p\times d}
\]
which reduces a vector $\x \in \R^p$ to its subvector $\bPi^\top\x = (x_i)_{i=1}^d$. Instead of looking for a suitable projection matrix $\A$ directly, we are looking for a suitable orthogonal matrix $\U$ such that $\hat{H}(\bPi^\top \U^\top \x_1, \ldots, \bPi^\top \U^\top \x_n)$ is particularly small.

For a rigorous description of the local search strategy, we need to introduce some notation and geometry. Recall first that any matrix space $\R^{p\times q}$ becomes a Euclidean space by means of the inner product $\langle \bs{B}, \bs{C}\rangle := \mathrm{trace}(\bs{B}^\top \bs{C}) = \sum_{i,j} B_{ij} C_{ij}$, and the resulting norm is the Frobenius norm $\|\bs{B}\|_F = \bigl( \sum_{i,j} B_{ij}^2 \bigr)^{1/2}$. In the special case of $p = q$, it is well-known that $\R^{p\times p}$ is the sum of the orthogonal linear spaces $\R^{p\times p}_{\rm sym}$ and $\R^{p\times p}_{\rm anti}$ of symmetric and antisymmetric matrices, respectively. Indeed, any matrix $\bDelta \in \R^{p\times p}$ can be written as $\bDelta = \bDelta_{\rm sym} + \bDelta_{\rm anti}$ with the symmetric matrix $\bDelta_{\rm sym} := 2^{-1} (\bDelta + \bDelta^\top)$ and the antisymmetric matrix $\bDelta_{\rm anti} := 2^{-1} (\bDelta - \bDelta^\top)$.

Searching locally means that a given candidate $\U$ in \eqref{eq:A.via.U} is multiplied from the right with another orthogonal matrix $\V$ which is close to the identity matrix $\bs{I}_p$. Specifically, let $\V$ be equal to
\[
	\exp(\bDelta) \ = \ \sum_{k=0}^\infty (k!)^{-1} \bDelta^k
\]
for $\bDelta \in \R^{p\times p}_{\rm anti}$. It is well-known that $\exp(\cdot)$ defines a surjective mapping from $\R^{p\times p}_{\rm anti}$ to the set of orthogonal matrices in $\R^{p\times p}$ with determinant $1$, where $\exp(\bs{0}) = \bs{I}_p$. Moreover, $\exp(\bDelta)^\top = \exp(-\bDelta)$, and
\[
	\exp(\bDelta) \ = \ \bs{I}_p + \bDelta + O(\|\bDelta\|_F^2) .
\]

To find a promising new orthogonal matrix $\U \exp(\bDelta)$, we may assume without loss of generality that $\U = \bs{I}_p$, because
\[
	\bPi^\top (\U \exp(\bDelta))^\top \x_i \ = \ \bPi^\top \exp(\bDelta)^\top (\U^\top \x_i) .	
\]
Hence, we may replace $\x_i$ with $\U^\top \x_i$ for $1 \le i \le p$ and then look for a promising perturbation $\exp(\bDelta)$ of $\bs{I}_p$. To this end, let
\begin{equation}
\label{eq:xi.in.blocks}
	\x_i \ = \ \begin{bmatrix} \y_i \\ \z_i \end{bmatrix}
	\quad\text{with} \ \y_i \in \R^d, \ \z_i \in \R^{p-d} ,
\end{equation}
that is, $\y_i = \bPi^\top \x_i$. If we write
\begin{equation}
\label{eq:Delta.in.blocks}
	\bDelta \ = \ \begin{bmatrix} \bDelta_1 & - \C^\top \\ \C & \bDelta_2 \end{bmatrix} 
\end{equation}
with arbitrary matrices $\bDelta_1 \in \R^{d\times d}_{\rm anti}$, $\bDelta_2 \in \R^{(p-d)\times (p-d)}_{\rm anti}$ and $\C \in \R^{(p-d)\times d}$, then
\[
	\bPi^\top \exp(\bDelta)^\top \x_i \
	= \ \y_i + \bDelta_1^\top \y_i + \C^\top \z_i + O(\|\bDelta\|_F^2) .
\]
Consequently, it follows from the general expansion of $\hat{H}(\cdot)$ in the previous section that
\begin{align}
\nonumber
	\hat{H} \bigl(
		&\bPi^\top \exp(\bDelta)^\top \x_1, \ldots, \bPi^\top \exp(\bDelta)^\top \x_n \bigr)
			- \hat{H}(\y_1,\ldots,\y_n) \\
\nonumber
	= \ & n^{-1} h^{-2} \sum_{i=1}^n
		\frac{\sum_{j=1}^n
			\bigl( \bDelta_1^\top (\y_i - \y_j)
				+ \C^\top (\z_i - \z_j) \bigr)^\top (\y_i - \y_j) \phi_j(\y_i - \y_j)}
			{\sum_{j=1}^n \phi_h(\y_i - \y_j)} \\
\nonumber
		& \qquad + \ O(\|\bDelta\|_F^2) \\
\label{eq:expansion.H}
	= \ & \langle \C, \hat{\C} \rangle + O(\|\bDelta\|_F^2)
\end{align}
as $\bDelta \to \bs{0}$, where
\begin{equation}
\label{eq:Chat}
	\hat{\C} \ := \ n^{-1} h^{-2} \sum_{i=1}^n
		\frac{\sum_{j=1}^n \phi_h(\y_i - \y_j) (\z_i - \z_j)(\y_i - \y_j)^\top}
			{\sum_{j=1}^n \phi_h(\y_i - \y_j)} \ \in \ \R^{(p-q)\times d} .
\end{equation}
In the last step we used the representations
\begin{align*}
	(\bDelta_1^\top (\y_i - \y_j))^\top (\y_i - \y_j) \
	&= \ \bigl\langle \bDelta_1, (\y_i - \y_j)(\y_i - \y_j)^\top \bigr\rangle , \\
	(\C^\top (\z_i - \z_j))^\top (\y_i - \y_j) \
	&= \ \bigl\langle \C, (\z_i - \z_j)(\y_i - \y_j)^\top \bigr\rangle ,
\end{align*}
and the fact that $\bDelta_1 \in \R^{d\times d}_{\rm anti}$ is perpendicular to $(\y_i - \y_j)(\y_i - \y_j)^\top \in \R^{d\times d}_{\rm sym}$. Since
\[
	\langle \C, \hat{\C}\rangle
	\ = \ 2^{-1} \langle \bDelta, \hat{\bDelta} \rangle
	\quad\text{with}\quad
	\hat{\bDelta} \ := \ \begin{bmatrix} \bs{0} & - \hat{\C}^\top \\ \hat{\C} & \bs{0} \end{bmatrix} ,
\]
expansion~\eqref{eq:expansion.H} shows that the gradient of the mapping
\[
	\R^{p\times p}_{\rm anti} \ni \bDelta
	\ \mapsto \ \hat{H} \bigl( \bPi^\top \exp(\bDelta)^\top \x_1, \ldots, \bPi^\top \exp(\bDelta)^\top \x_n \bigr)
\]
at $\bDelta = \bs{0}$ is given by $2^{-1} \hat{\bDelta}$. Consequently, promising candidates for the factor $\exp(\bDelta)$ are given by
\[
	\exp(- t \hat{\bDelta}) = \exp(t \hat{\bDelta})^\top ,
	\quad t \ge 0 .
\]
The explicit computation of $\exp(t\hat{\bDelta})$ is rather convenient when working with a singular value decomposition of $\hat{\C}$. With $m := \min(d,p-d)$, suppose that
\[
	\hat{\C} \ = \ \W \diag(\bs{\sigma}) \V^\top
\]
with matrices $\W = [\w_1 \cdots \w_m] \in \R^{(p-d)\times m}$, $\V = [\v_1 \cdots \v_m] \in \R^{d\times m}$ of singular vectors such that $\W^\top\W = \V^\top\V = \bs{I}_m$ and a vector $\bs{\sigma} \in [0,\infty)^m$ of singular values. If we define
\[
	\hat{\v}_i := \begin{bmatrix} \v_i \\ \bs{0} \end{bmatrix} \in \R^p
	\quad\text{and}\quad
	\hat{\w}_i := \begin{bmatrix} \bs{0} \\ \w_i \end{bmatrix} \in \R^p ,
\]
then $\hat{\bDelta} \hat{\v}_i = \sigma_i \hat{\w}_i$ and $\hat{\bDelta} \hat{\w}_i = - \sigma_i \hat{\v}_i$ for $1 \le i \le m$, while $\hat{\bDelta}\x = \bs{0}$ whenever $\x \perp \{\hat{\v}_1,\ldots,\hat{\v}_m,\hat{\w}_1,\ldots,\hat{\w}_m\}$. From this one can deduce that for $1 \le i \le m$,
\begin{align*}
	\exp(t\hat{\bDelta}) \hat{\v}_i \
	&= \quad \cos(t\sigma_i) \hat{\v}_i + \sin(t\sigma_i) \hat{\w}_i , \\
	\exp(t\hat{\bDelta}) \hat{\w}_i \
	&= \ - \sin(t\sigma_i) \hat{\v}_i + \cos(t\sigma_i) \hat{\w}_i ,
\end{align*}
while $\exp(t\hat{\bDelta}) \x = \x$ for $\x \perp \{\hat{\v}_1,\ldots,\hat{\v}_m,\hat{\w}_1,\ldots,\hat{\w}_m\}$. In other words,
\begin{align*}
	\exp(t \hat{\bDelta}) \
	&= \ \begin{bmatrix}
		\bs{I}_d + \V \diag(\cos(t\bs{\sigma}) - 1) \V^\top
			& - \V \diag(\sin(t\bs{\sigma})) \W^\top \\
		\W \diag(\sin(t\bs{\sigma})) \V^\top
			& \bs{I}_{p-d} + \W \diag(\cos(t\bs{\sigma}) - 1) \W^\top
		\end{bmatrix} \\
	&= \ \begin{bmatrix}
		\bs{I}_d - \V \diag(2 \sin(t\bs{\sigma}/2)^2) \V^\top
			& - \V \diag(\sin(t\bs{\sigma})) \W^\top \\
		\W \diag(\sin(t\bs{\sigma})) \V^\top
			& \bs{I}_{p-d} - \W \diag(2 \sin(t\bs{\sigma}/2)^2) \W^\top
		\end{bmatrix} ,
\end{align*}
where the functions of $t \bs{\sigma}$ are computed component-wise. Note that the upper left block $\bs{I}_d + \V \diag(\cos(t\bs{\sigma}) - 1) \V^\top$ equals $\V \diag(\cos(t\bs{\sigma})) \V^\top$ in case of $d = m$, and the lower right block $\bs{I}_{p-d} + \W \diag(\cos(t\bs{\sigma}) - 1) \W^\top$ equals $\W \diag(\cos(t\bs{\sigma})) \W^\top$ in case of $p-d = m$.

For the explicit choice of $t \ge 0$, we propose an Armijo--Goldstein procedure, see \cite{Nocedal_Wright_2006}. Specifically, recall that the auxiliary function
\[
	\hat{h}(t) \
	:= \ \hat{H} \bigl( \bPi^\top \exp(t\hat{\bDelta}) \x_1, \ldots,
		\bPi^\top \exp(t\hat{\bDelta}) \x_n \bigr)
\]
satisfies $\hat{h}(0) = \hat{H}(\y_1,\ldots,\y_n)$ and $\hat{h}'(0) = - \|\hat{\C}\|_F^2$. Now we choose $t = 2^{-k}$ with the smallest integer $k \ge 0$ such that the improvement $\hat{h}(0) - \hat{h}(2^{-k})$ is at least $- 2^{-k} \hat{h}'(0)/3 = 2^{-k} \|\hat{\C}\|_F^2/3$.

\paragraph{Using arbitrary PP indices.}
Suppose we replace estimated entropy with an arbitrary continuously differentiable function $\hat{H} : (\R^d)^n \to \R$. Then there exist vectors $\bs{\gamma}_i = \bs{\gamma}_i(\y_1,\ldots,\y_n) \in \R^d$ such that for arbitrary perturbations $\bs{\delta}_i \in \R^d$,
\[
	\hat{H}(\y_1 + \bs{\delta}_1,\ldots,\y_n+ \bs{\delta}_n) \
	= \ \hat{H}(\y_1,\ldots,\y_n) + \sum_{i=1}^n \bs{\gamma}_i^\top \bs{\delta}_i^{}
		+ o \Bigl( \sum_{i=1}^n \|\bs{\delta}_i\| \Bigr)
\]
as $\bs{\delta}_1,\ldots,\bs{\delta}_n \to \bs{0}$. With $\x_i$ and $\bDelta$ as in \eqref{eq:xi.in.blocks} and \eqref{eq:Delta.in.blocks},
\begin{align*}
	\hat{H} &\bigl(
		\bPi^\top \exp(\bDelta)^\top \x_1, \ldots, \bPi^\top \exp(\bDelta)^\top \x_n \bigr) \\
	&= \ \hat{H}(\y_1,\ldots,\y_n)
		+ \sum_{i=1}^n \bs{\gamma}_i^\top (\bDelta_1^\top \y_i + \C^\top \z_i)
		+ o(\|\bDelta\|_F)
\end{align*}
as $\bDelta \to \bs{0}$. If $\hat{H}$ is orthogonally invariant in the sense that $\hat{H}(\bs{V}\y_1,\ldots,\bs{V}\y_n) = \hat{H}(\y_1,\ldots,\y_n)$ for arbitrary orthogonal matrices $\bs{V} \in \R^{d\times d}$, then $\sum_{i=1}^n \bs{\gamma}_i^\top \bDelta_1^\top \y_i = 0$. This can be seen by considering the special case $\C = \bs{0}$ and $\bDelta_2 = \bs{0}$, because then $\bPi^\top \exp(\bDelta)^\top \x_i = \exp(\bDelta_1)^\top \y_i$, and $\exp(\bDelta_1)$ is orthogonal. Consequently, the previous expansion of $\hat{H}$ simplifies to
\[
	\hat{H} \bigl(
		\bPi^\top \exp(\bDelta)^\top \x_1, \ldots, \bPi^\top \exp(\bDelta)^\top \x_n \bigr) \\
	= \ \hat{H}(\y_1,\ldots,\y_n) + \langle \C, \hat{\C} \rangle + o(\|\bDelta\|_F)
\]
as $\bDelta \to \bs{0}$, where
\[
	\hat{\C} \ := \ \sum_{i=1}^n \z_i^{} \bs{\gamma}_i^\top \ \in \R^{(p-d)\times d} .
\]
Hence, our version of local optimization may be applied to any smooth PP index $\hat{H}$ which is orthogonally invariant.

\section{The complete procedure(s)}

The complete procedure consists of three different basic procedures.

\paragraph{Basic procedure 1 (Pre-whitening).}
Given the centered raw data $\x_1^{\rm raw}, \ldots, \x_n^{\rm raw}$, we compute the preliminary scatter estimator $\hat{\Sigma}_0(\x_1^{\rm raw}, \ldots, \x_n^{\rm raw}) = \bs{B}_0^{} \bs{B}_0^\top$. Then we set
\[
	\x_i^{\rm pre} \ := \ \bs{B}_0^{-1} \x_i^{\rm raw} .
\]

\paragraph{Basic procedure 2 (ICS).}
Now we compute the second scatter estimator and its spectral decomposition: $\hat{\Sigma}(\x_1^{\rm pre},\ldots,\x_n^{\rm pre}) = \hat{\U} \diag(\hat{\bs{\lambda}}) \hat{\U}^\top$ with an orthogonal matrix $\hat{\U} \in \R^{p\times p}$ and a vector $\hat{\bs{\lambda}} \in (0,\infty)^p$ of eigenvectors. Then we set
\[
	\x_i^{\rm ics} \ := \ \hat{\U}^\top \x_i^{\rm pre} .
\]

\paragraph{Basic procedure 3 (Local PP).}
For given indices $j(1) < \cdots < j(d)$ in $\{1,2,\ldots,p\}$, let $\ell(1) < \cdots < \ell(p-d)$ be the elements of $\{1,2,\ldots,p\} \setminus \{j(1),\ldots,j(d)\}$. With the standard basis $\bs{e}_1, \bs{e}_2, \ldots, \bs{e}_p$ of $\R^p$, we define the permutation matrix $\U := [\bs{e}_{j(1)} \cdots \bs{e}_{j(d)} \, \bs{e}_{\ell(1)} \cdots \bs{e}_{\ell(p-d)}]$ and set
\[
	(\x_1,\ldots,\x_n) \ \leftarrow \ (\U^\top \x_1^{\rm ics}, \ldots, \U^\top \x_n^{\rm ics}) .
\]
Now we start the following iterative algorithm with some small threshold $\delta_o > 0$:
\[
	\begin{array}{|l|}
	\hline
	\hat{H} \leftarrow \hat{H}(\bPi^\top \x_1, \ldots, \bPi^\top \x_n)^{\strut} \\
	\hat{\C} \leftarrow \hat{\C}(\x_1,\ldots,\x_n) \\
	\delta \leftarrow \|\hat{\C}\|_F^2 \\
	\text{while} \ \delta \ge \delta_o \ \text{do}\\
	\qquad (\V,\bs{\sigma},\W) \leftarrow \mathrm{SVD}(\hat{\C}) \\
	\qquad \U \leftarrow \mathrm{Exp}(\V,\bs{\sigma},\W) \\
	\qquad (\x_1^{\rm tmp},\ldots,\x_n^{\rm tmp}) \leftarrow (\U\x_1,\ldots,\U\x_n) \\
	\qquad \hat{H}^{\rm tmp} \leftarrow
		\hat{H}(\bPi^\top \x_1^{\rm tmp},\ldots,\bPi^\top \x_n^{\rm tmp}) \\
	\qquad \text{while} \ \hat{H} - \hat{H}^{\rm tmp} < \delta/3 \ \text{do} \\
	\qquad\qquad \delta \leftarrow \delta/2 \\
	\qquad\qquad \bs{\sigma} \leftarrow \bs{\sigma}/2 \\
	\qquad\qquad \U \leftarrow \mathrm{Exp}(\V,\bs{\sigma},\W) \\
	\qquad\qquad (\x_1^{\rm tmp},\ldots,\x_n^{\rm tmp}) \leftarrow (\U\x_1,\ldots,\U\x_n) \\
	\qquad\qquad \hat{H}^{\rm tmp} \leftarrow
		\hat{H}(\bPi^\top \x_1^{\rm tmp},\ldots,\bPi^\top \x_n^{\rm tmp}) \\
	\qquad \text{end while} \\
	\qquad (\x_1,\ldots,\x_n) \leftarrow (\x_1^{\rm tmp},\ldots,\x_n^{\rm tmp}) \\
	\qquad \hat{H} \leftarrow \hat{H}^{\rm tmp} \\
	\qquad \hat{\C} \leftarrow \hat{\C}(\x_1,\ldots,\x_n) \\
	\qquad \delta \leftarrow \|\hat{\C}\|_F^2 \\
	\text{end while}\\
	\hline
	\end{array}
\]
Here $\hat{\C}(\x_1,\ldots,\x_n) \in \R^{(p-d)\times d}$ is given by \eqref{eq:Chat}, $\mathrm{SVD}(\hat{\C})$ yields the ingredients for the singular value decomposition $\hat{\C} = \W \diag(\bs{\sigma}) \V^\top$, and $\mathrm{Exp}(\V,\bs{\sigma},\W)$ computes
\[
	\exp \biggl(\begin{bmatrix}
		\bs{0} & - \V \diag(\bs{\sigma}) \W^\top \\
		\W \diag(\bs{\sigma}) \V^\top & \bs{0}
	\end{bmatrix} \biggr) .
\]
The inner while-loop is the Armijo--Goldstein step size correction mentioned before.

The iterative algorithm will always converge. In each instance of the outer while-loop, the data $(\x_1,\ldots,\x_n)$ is replaced with $(\U\x_1,\ldots,\U\x_n)$ with some orthogonal matrix $\U$ such that $\hat{H}(\bPi^\top \x_1, \ldots, \bPi^\top \x_n)$ decreases strictly. Since the set of all orthogonal matrices is a compact, differentiable manifold, and since $\hat{\C}(\x_1,\ldots,\x_n)$ is a continuous function of its input data which is closely related to the gradient of $\hat{H}(\bPi^\top \U\x_1, \ldots, \bPi^\top \U\x_n)$ as a function of $\U$, the condition $\|\hat{\C}\|_F^2 < \delta_o$ has to be satisfied after finitely many steps.

Basic procedure~3 is executed with $d+1$ or $\binom{p}{d}$ different choices of $i(1) < \cdots < i(d)$. It is also possible to compute first $\hat{H}(\bPi^\top \x_1, \ldots, \bPi^\top \x_n)$ for all these choices and then start local PP only for the choice with minimal initial value of $\hat{H}$. Alternatively, one can inspect a scatter plot matrix of the data $\x_i^{\rm ics}$ visually and then run a local search, either with $d = 1$ and the most promising index $j$, or with $d = 2$ and the most promising indices $1 \le j(1) <  j(2) \le p$.

\paragraph{The result.}
Running basic procedures 1, 2 and 3 leads to transformed observations $\x_i = \bs{B}^\top \x_i^{\rm raw}$ such that the $d$-dimensional observations $\bPi^\top\x_i$ reveal (hopefully) some interesting feature of the raw data.

The nonsingular matrix $\bs{B} \in \R^{p\times p}$ may be recovered quickly via multivariate least squares: With the data matrices $\underline{\bs{X}}_{\rm raw} = [\x_1^{\rm raw} \, \ldots \,\x_n^{\rm raw}]^\top$ and $\underline{\bs{X}} = [\x_1 \, \ldots \, \x_n]^\top$ in $\R^{n\times p}$, the matrix $\bs{B}$ satisfies $\underline{\bs{X}}_{\rm raw} \bs{B} = \underline{\bs{X}}$, whence $\bs{B} = (\underline{\bs{X}}_{\rm raw}^\top \underline{\bs{X}}_{\rm raw}^{})^{-1} \underline{\bs{X}}_{\rm raw}^\top \underline{\bs{X}}$.

We did not specify how the raw data have been centered. In our numerical experiments we used the sample mean, but any estimator of location would be possible, provided that the scatter estimators $\hat{\Sigma}_0(\x_1^{\rm raw}, \ldots, \x_n^{\rm raw})$ and $\hat{\Sigma}(\x_1^{\rm pre}, \ldots, \x_n^{\rm pre})$ are invariant under translations of the input data. Note that $\hat{H}$ has this invariance property as well.

\paragraph{Global PP.}
In our numerical experiments, we also tried a global version of PP. This consists of basic procedure~1 (prewhitening) and basic procedure~3 (local PP) applied to the observations $\x_i^{\rm pre}$ instead of the observations $\x_i^{\rm ics}$. Of course, one could extend this by applying basic procedure~3 several times to the observations $\V_s^\top \x_i^{\rm pre}$, where $\V_1, \V_2, \V_3, \ldots$ are independent random orthogonal matrices in $\R^{p\times p}$, independent from the data.

\section{Numerical examples}

The subsequent numerical examples are similar to examples presented by \cite{Tyler_etal_2009}, but with higher dimensions. We always started with the sample covariance matrix $\hat{\bSigma}_0$, and $\hat{\bSigma}(\x_1,\ldots,\x_n)$ was a one-step symmetrized $M$-estimator of scatter,
\begin{equation}
\label{eq:one-step-M}
	\hat{\bSigma}(\x_1,\ldots,\x_n) \ := \ C \sum_{1 \le i < j \le n}
		\frac{(\x_i - \x_j)(\x_i - \x_j)^\top}{(\nu + \|\x_i - \x_j\|^2)_{}^\gamma}
\end{equation}
with some (irrelevant for us) scaling factor $C = C_{n,p,\nu,\gamma} > 0$ and parameters $\nu, \gamma > 0$. If $\nu > 0$, $\gamma = 1$ and $C = \nu + p$, then $\hat{\bSigma}$ is a one-step approximation of the symmetrized maximum-likelihood estimator of a centered multivariate $t$ distribution with $\nu$ degrees of freedom, see \cite{Kent_Tyler_1991}. If $\nu = 0$ and $\gamma = 1$, then $\hat{\bSigma}$ corresponds to the symmetrized distribution-free scatter estimator of \cite{Tyler_1987}. Our numerical experiments indicate that it is worthwhile to try $\nu$ close to $0$ and various parameters $\gamma > 1$, although $\hat{\bSigma}$ does not correspond to a robust $M$-estimator of scatter then.

Another question is the choice of the bandwidth $h > 0$. The larger the bandwidth $h$, the smoother is the target function $\hat{H}$, whereas small bandwiths $h$ result in many irrelevant local minima of $\hat{H}$. Our numerical experiments indicate that to detect clusters, values $h$ between $0.3$ and $0.5$ work quite well. However, if the structure to be detected is on a rather small scale, e.g.\ data lying on several parallel hyperplanes, then one needs smaller bandwidths $h$ (and probably exponents $\gamma > 1$) to detect such features.

For all subsequent examples, we simulated data sets of size $n = 500$ in different dimensions $p$, and we searched for interesting projections in dimension $d=2$. The underlying distribution was chosen such that a scatter plot of the raw or prewhitened data would not reveal the interesting structure. As to basic procedure~3, we considered all $p(p-1)/2$ standard projections $\x \mapsto \bPi^\top\U_{jk}^\top\x$, where $1 \le j < k \le p$ and $\U_{jk} = [\bs{e}_j, \bs{e}_k, \bs{e}_{\ell(1)}, \ldots, \bs{e}_{\ell(p-2)}]$ with the elements $\ell(1) < \cdots < \ell(p-2)$ of $\{1,\ldots,p\} \setminus \{j,k\}$. Local PP was run with threshold $\delta_o = 10^{-11}$.

\begin{figure}
\includegraphics[width=0.49\textwidth]{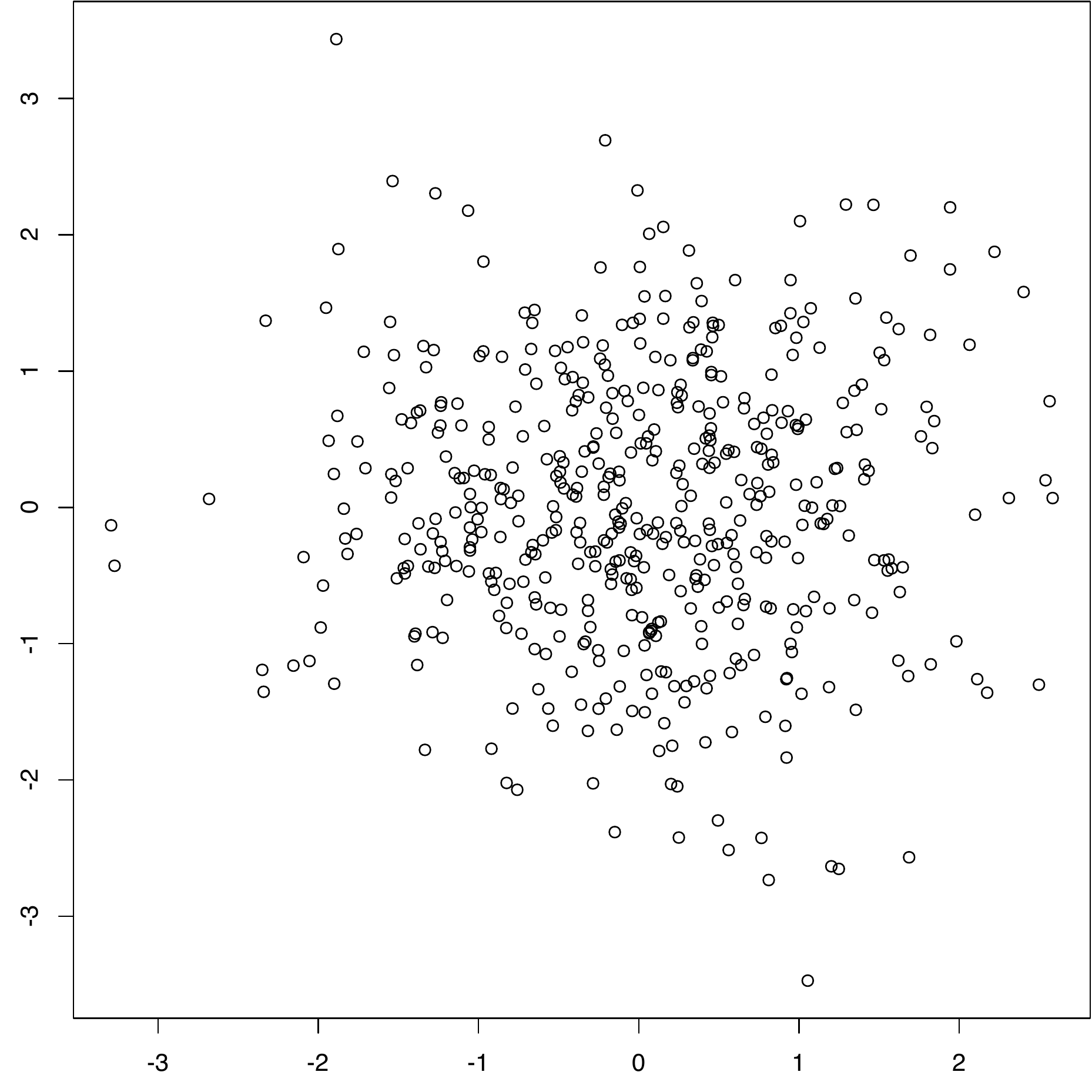}
\hfill
\includegraphics[width=0.49\textwidth]{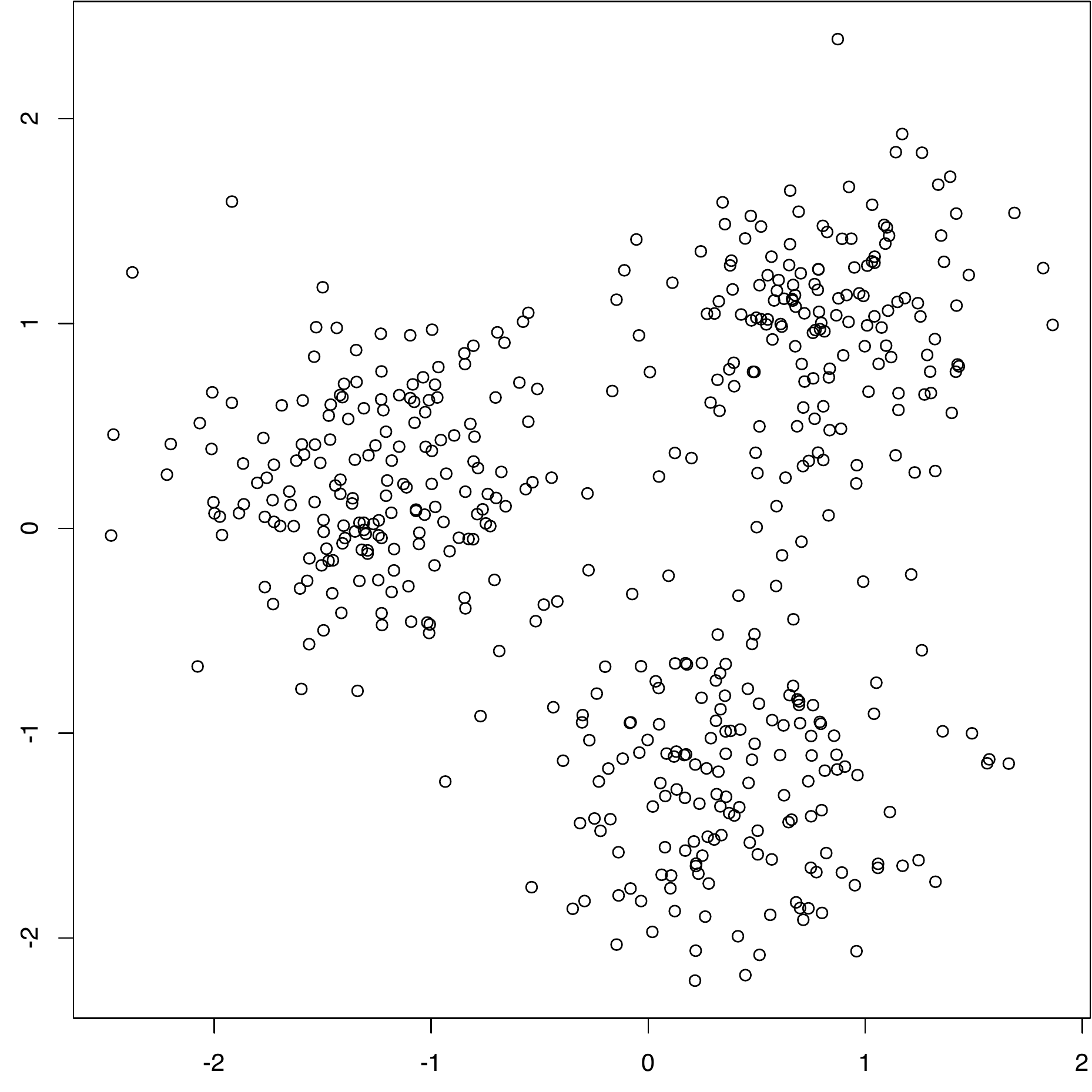}

\includegraphics[width=0.49\textwidth]{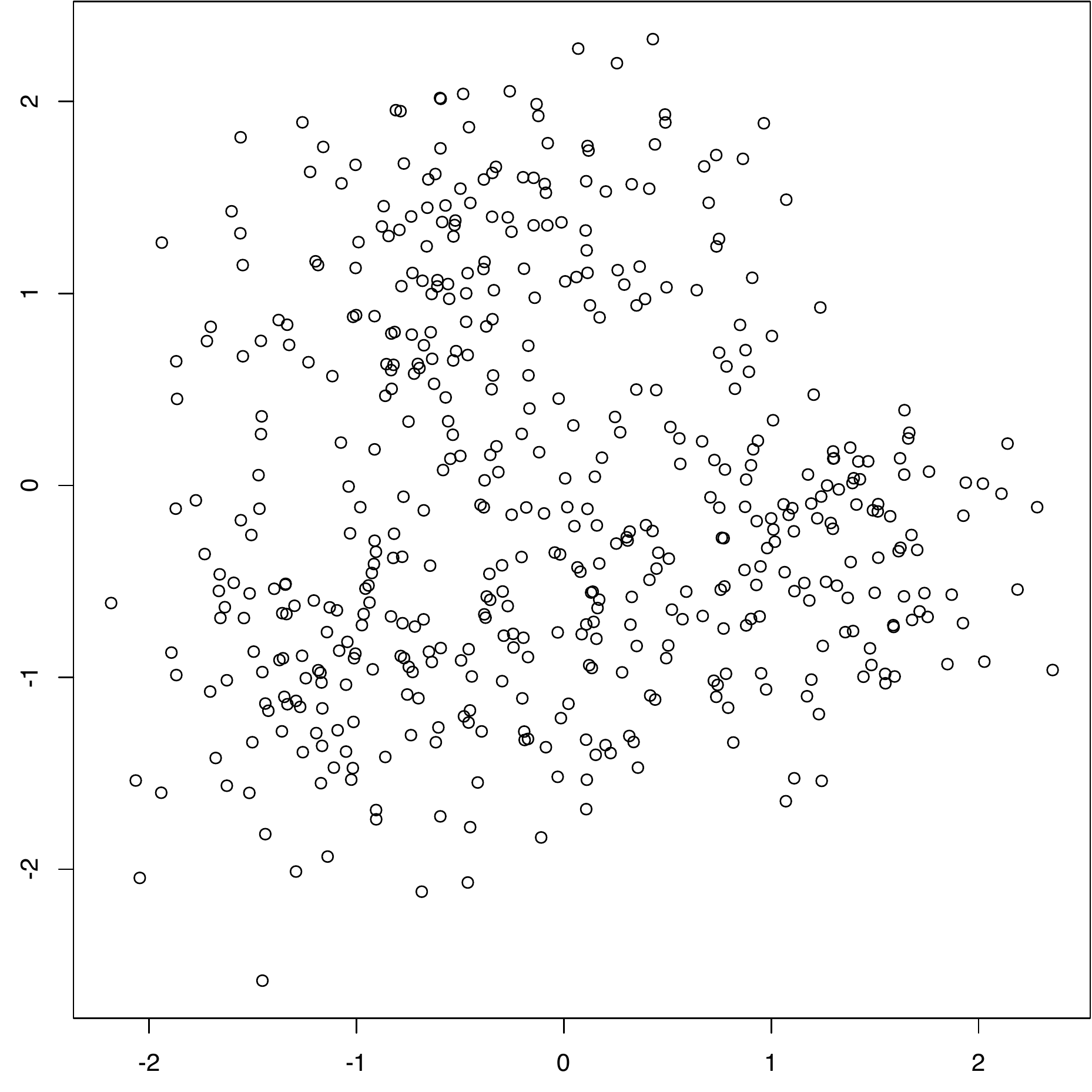}
\hfill
\includegraphics[width=0.49\textwidth]{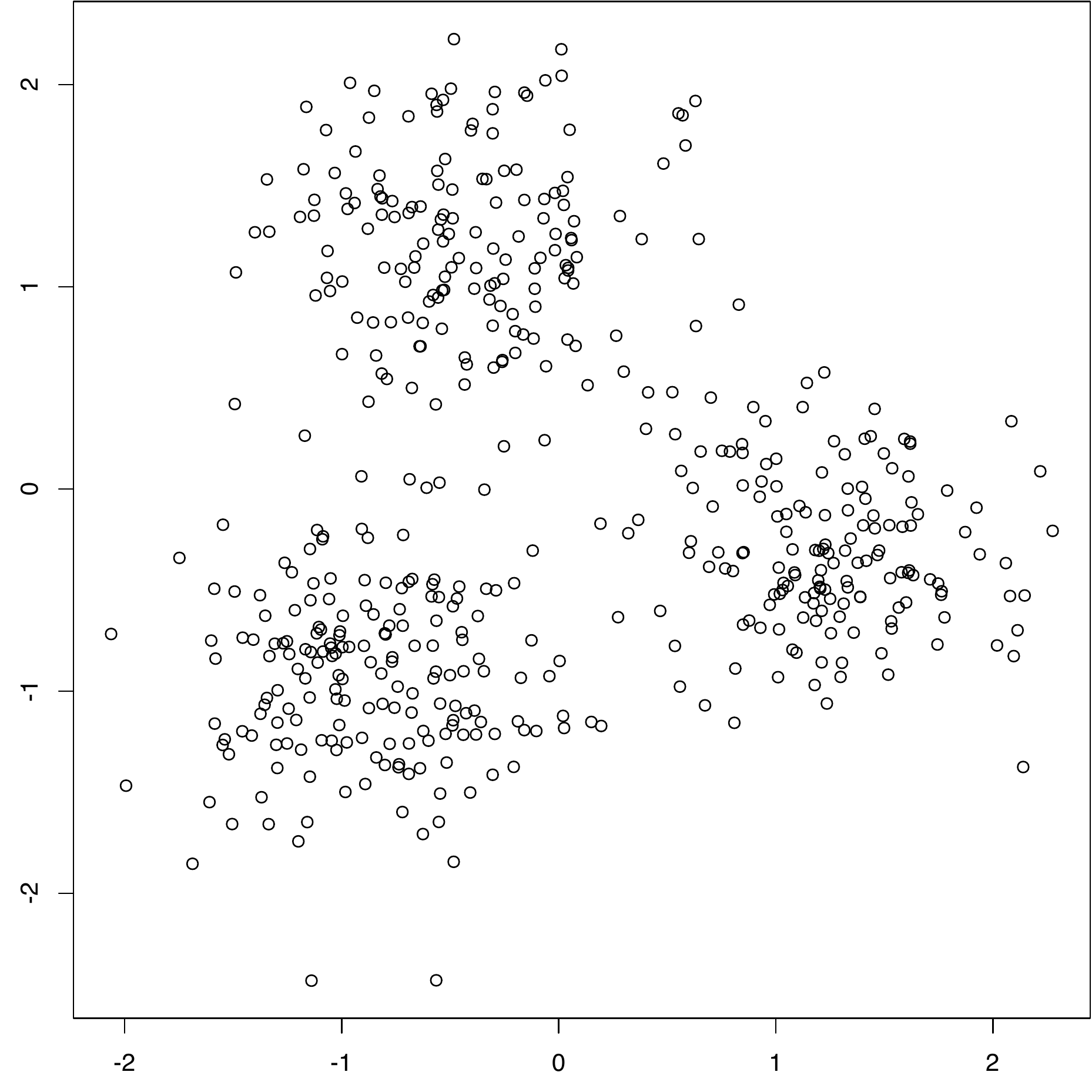}

\caption{PP for Example~1: Two-dimensional projections of the prewhitened data $\x_i$ before and after local PP (upper left and right panel), and of the preprocessed data $\x_i^{\rm ics}$ before and after local PP (lower left and right panel).}
\label{fig:Ex1}
\end{figure}

\paragraph{Example~1.}
We simulated data in dimension $p = 8$. After pre-whitening the data, we first tried a global PP with $h = 0.5$, which yields a value of $2.8610$ for \eqref{eq:standard.value.H}. The initial values $\hat{H}(\bPi^\top \U_{jk}^\top \x_1^{\rm pre}, \ldots, \bPi^\top\U_{jk}^\top \x_n^{\rm pre})$ ranged from $2.8251$ to $2.8423$. The minimal value was obtained for $(j,k) = (1,5)$. Running a local PP with this starting point revealed three clusters, and the final value of $\hat{H}$ was $2.4735$. This is shown in the upper panels of Figure~\ref{fig:Ex1}; the scatter plots show the projections before (left) and after (right) local PP.

Now we applied the procedure we advocate in this manuscript. We performed ICS with $\nu = 0$ and $\gamma = 1$. Then the values $\hat{H}(\bPi^\top \U_{jk}^\top \x_1^{\rm ics}, \ldots, \bPi^\top\U_{jk}^\top \x_n^{\rm ics})$ ranged from $2.7112$ to $2.8395$. The minimal value was obtained for $(j,k) = (1,2)$, and the corresponding scatter plot indicates already some clustering, see the lower left panel of Figure~\ref{fig:Ex1}. Running local PP revealed essentially the same structure as the global PP, see the lower right panel.

In this example, global PP without ICS seems to be just as good as our three-stage procedure. But note that local PP starting from components $1$ and $5$ of the data $\x_i^{\rm pre}$ resulted in $24$ iterations, whereas local PP starting from components $1$ and $2$ of the data $\x_i^{\rm ics}$ resulted in $10$ iterations only.

\begin{figure}
\includegraphics[width=0.49\textwidth]{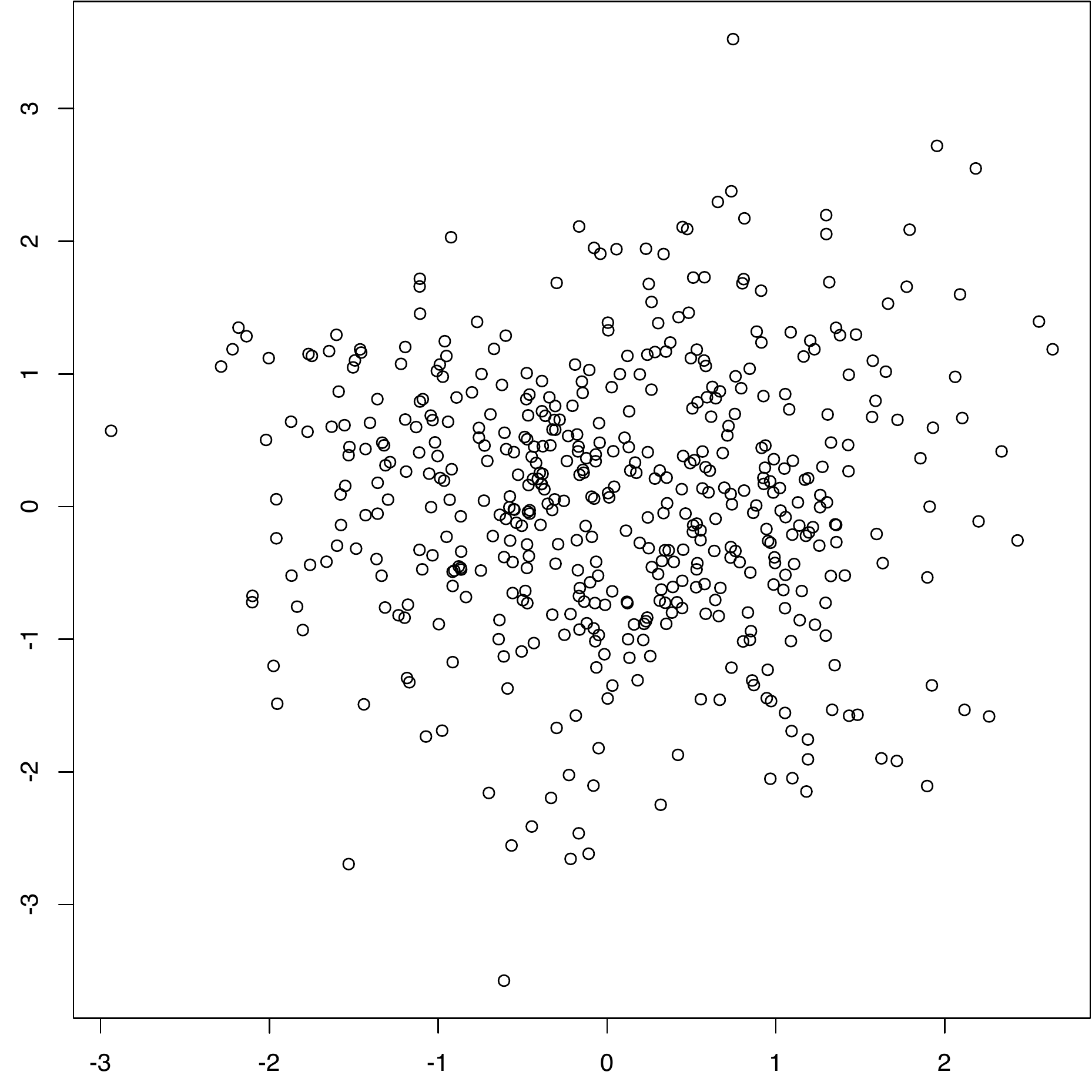}
\hfill
\includegraphics[width=0.49\textwidth]{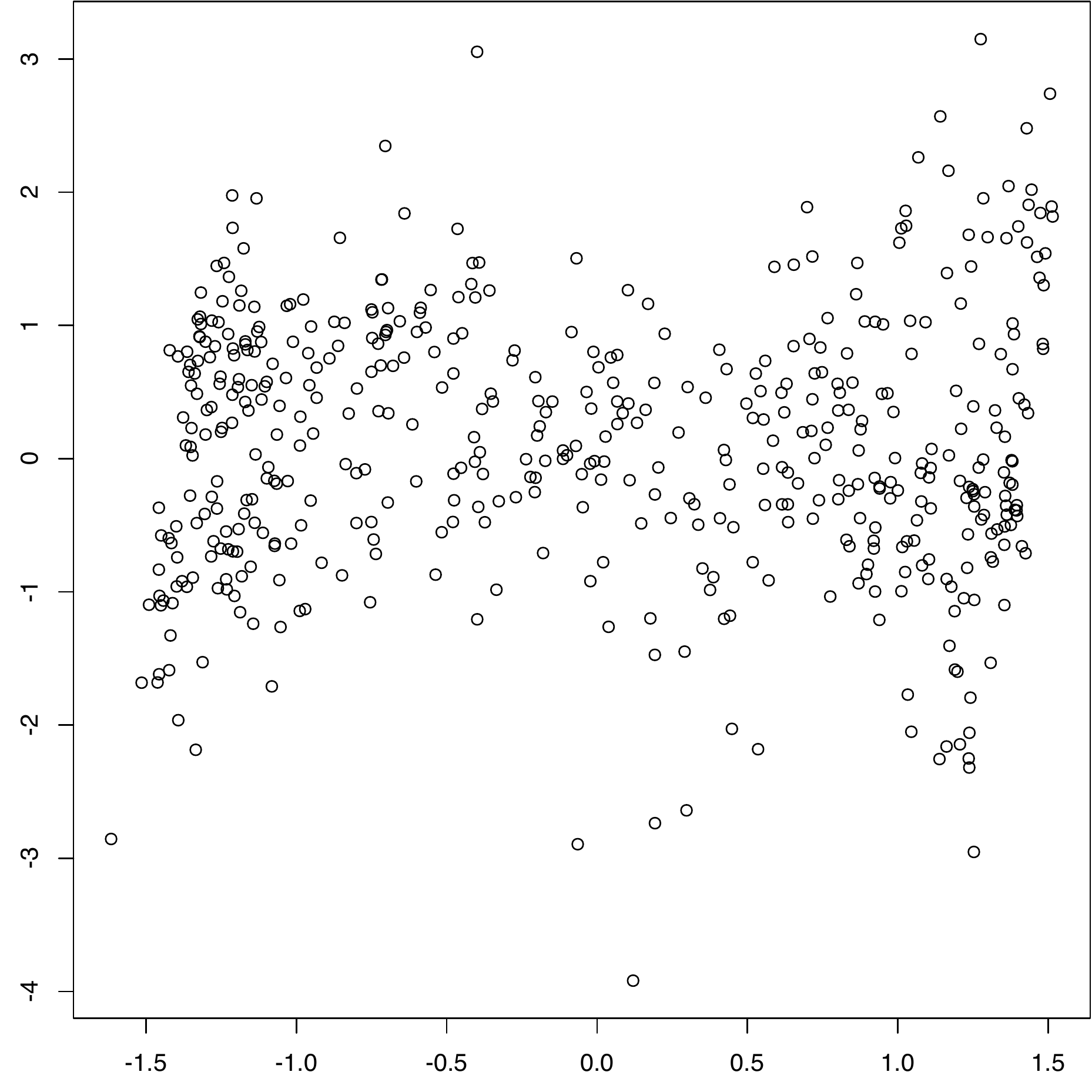}

\includegraphics[width=0.49\textwidth]{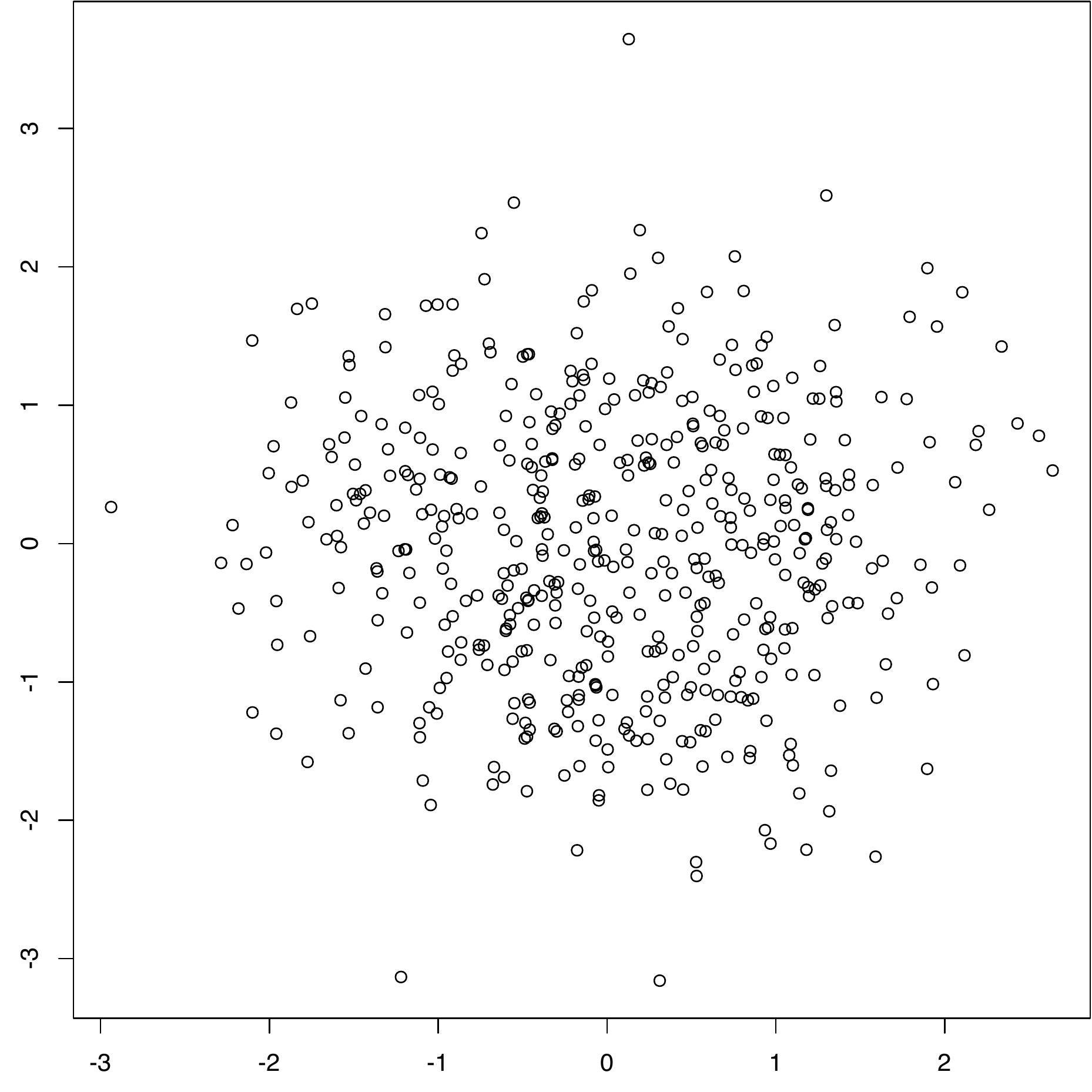}
\hfill
\includegraphics[width=0.49\textwidth]{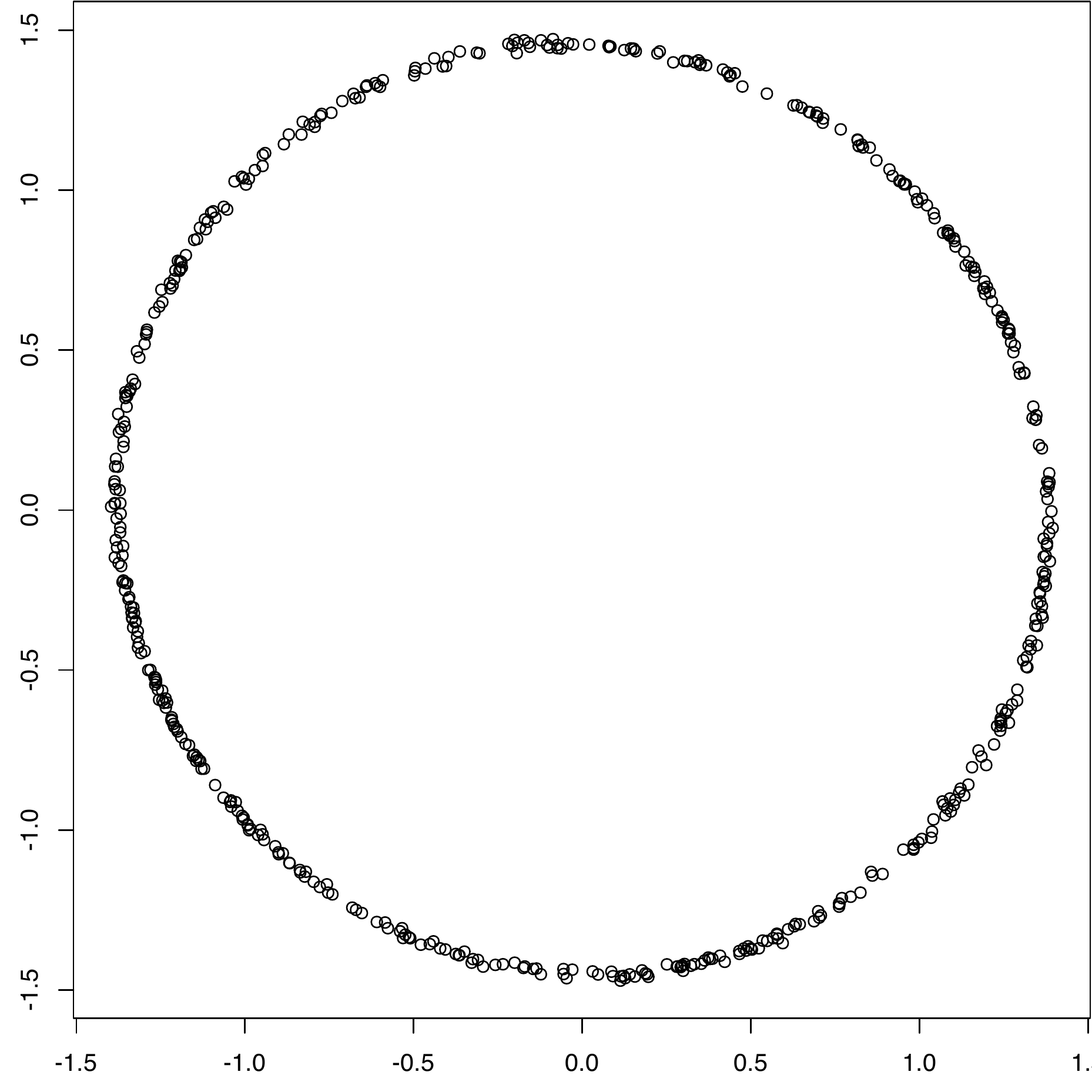}

\caption{PP for Example~2: Two-dimensional projections of the data $\x_i^{\rm pre}$ before (left panels) and after (right panels) local PP. The upper row corresponds to components $(j,k) = (3,15)$, the lower row to components $(j,k)=(3,6)$.}
\label{fig:Ex2A}
\end{figure}

\paragraph{Example~2.}
We simulated data in dimension $p = 16$. Again, we tried first a global PP without ICS, that means, we started local PP from some standard projections of the data $\x_i^{\rm pre}$. With the same bandwidth $h = 0.5$ as in Example~1, the initial values
\[
	\hat{H}_{jk}
	\ := \ \hat{H}(\bPi^\top \U_{jk}^\top \x_1^{\rm pre}, \ldots,
		\bPi^\top\U_{jk}^\top \x_n^{\rm pre})
\]
ranged from $2.8133$ to $2.8472$. The minimal value was obtained for $(j,k) = (3,15)$. The upper left panel of Figure~\ref{fig:Ex2A} shows that projection of the data $\x_i^{\rm pre}$, and starting local PP from this projection led to the scatter plot in the upper right panel with a value $2.6599$ of $\hat{H}$. The number of iterations was $112$.

Next, we tried other index pairs $(j,k)$, ordered by the initial values $\hat{H}_{jk}$. The detected structures were similar for the next $12$ pairs, but starting local PP from $(j,k) = (3,6)$ revealed the true underlying structure, a uniform distribution on a two-dimensional circle with a value $2.3871$ of $\hat{H}$, see the lower panels of Figure~\ref{fig:Ex2A}. The number of iterations was $29$.

\begin{figure}
\includegraphics[width=0.49\textwidth]{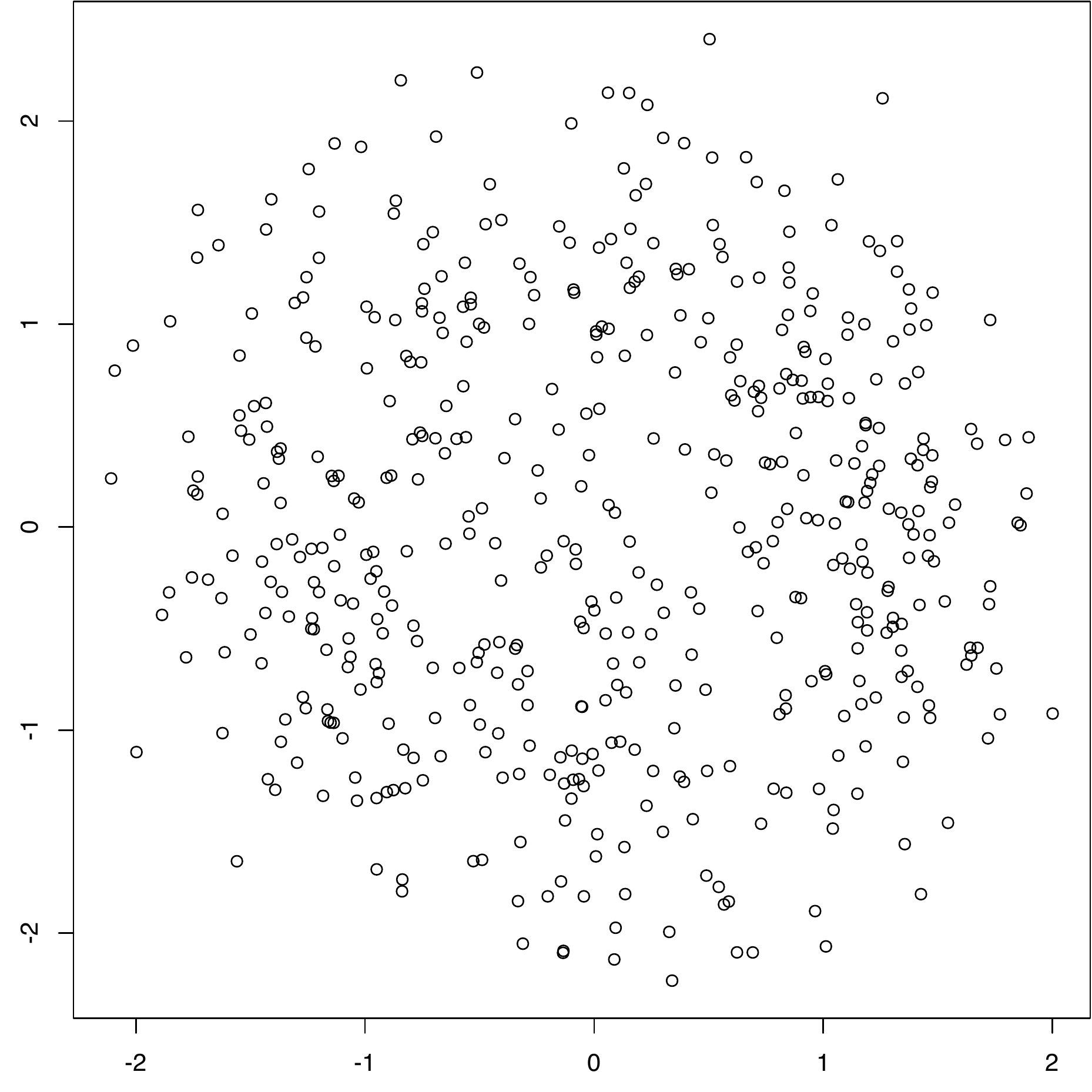}
\hfill
\includegraphics[width=0.49\textwidth]{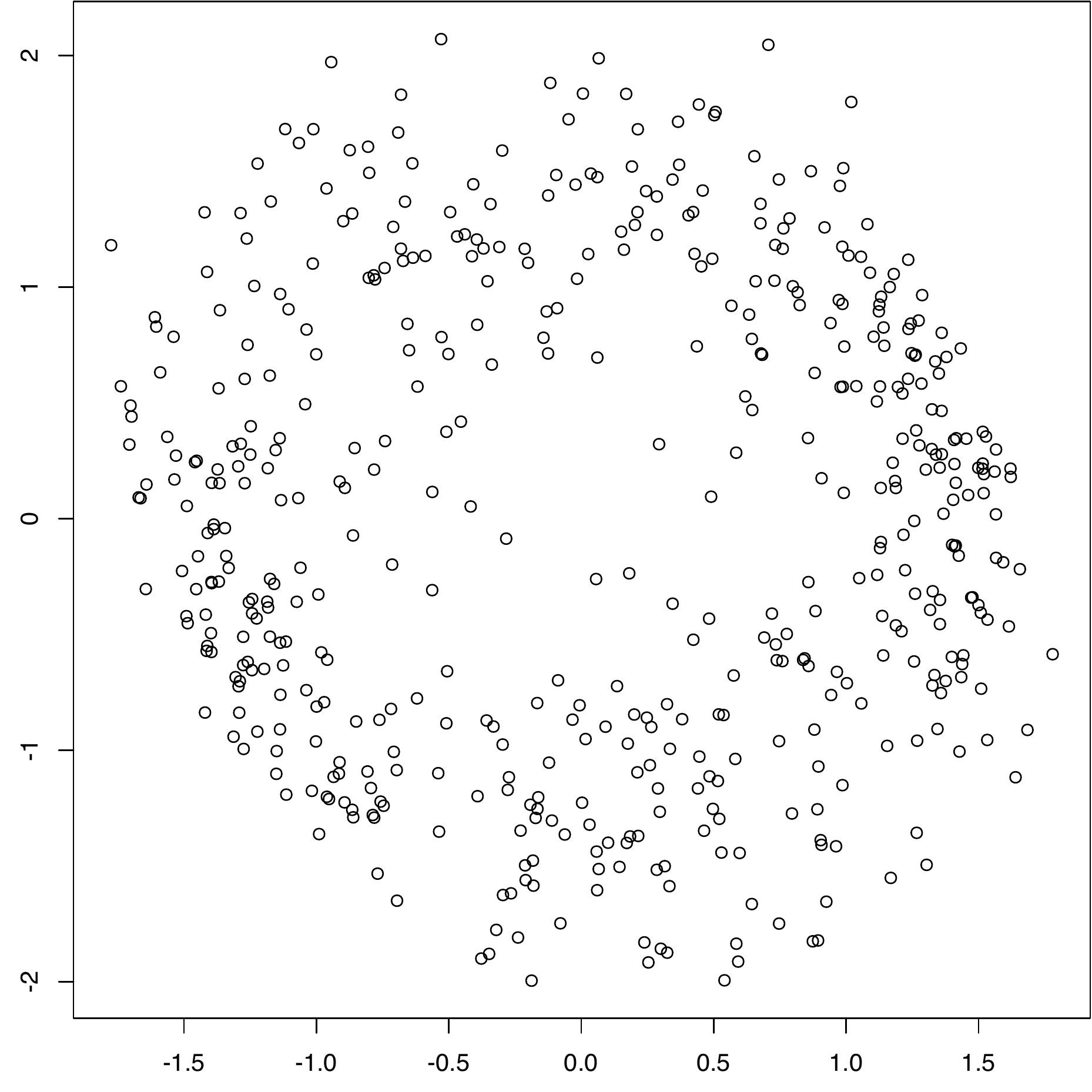}

\includegraphics[width=0.49\textwidth]{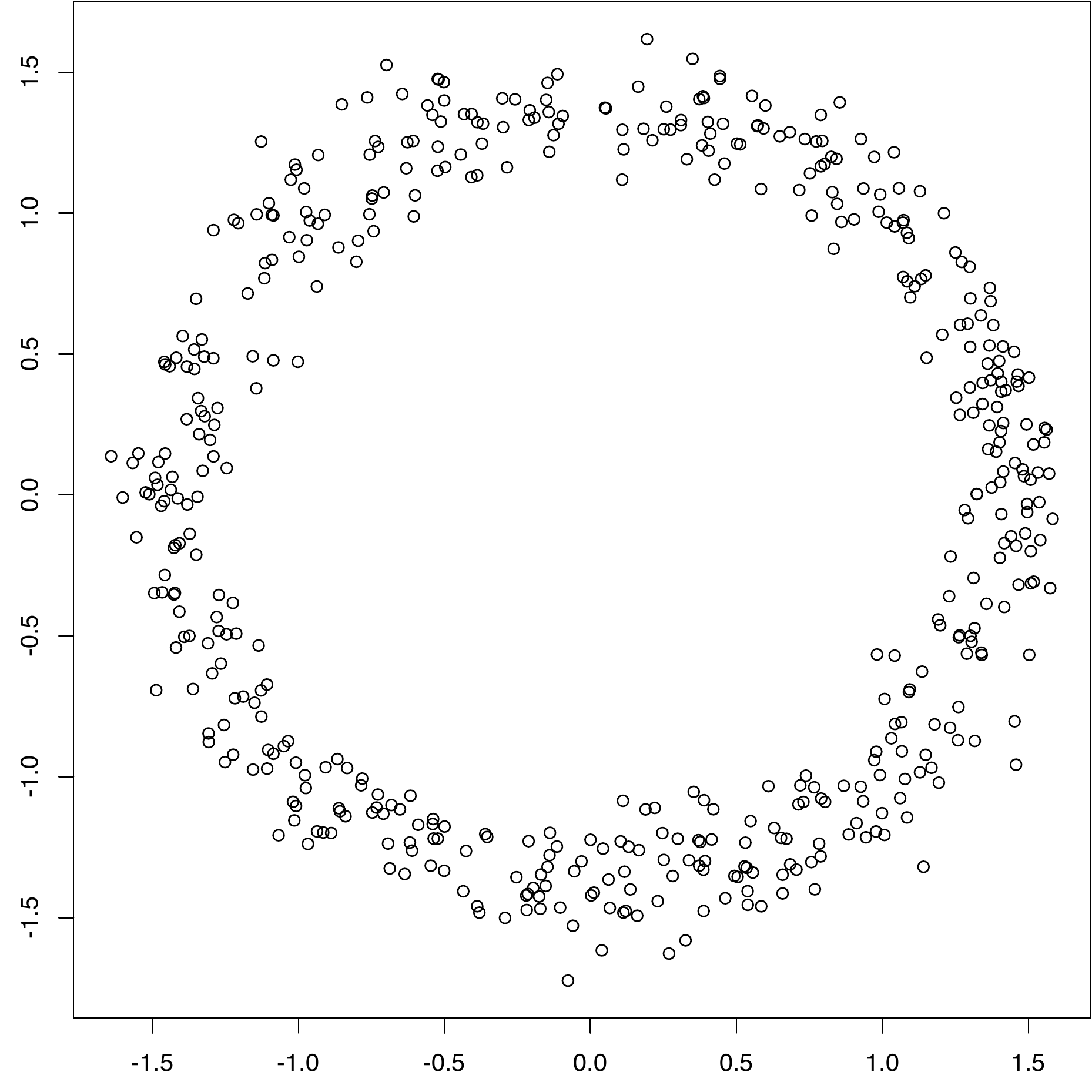}
\hfill
\includegraphics[width=0.49\textwidth]{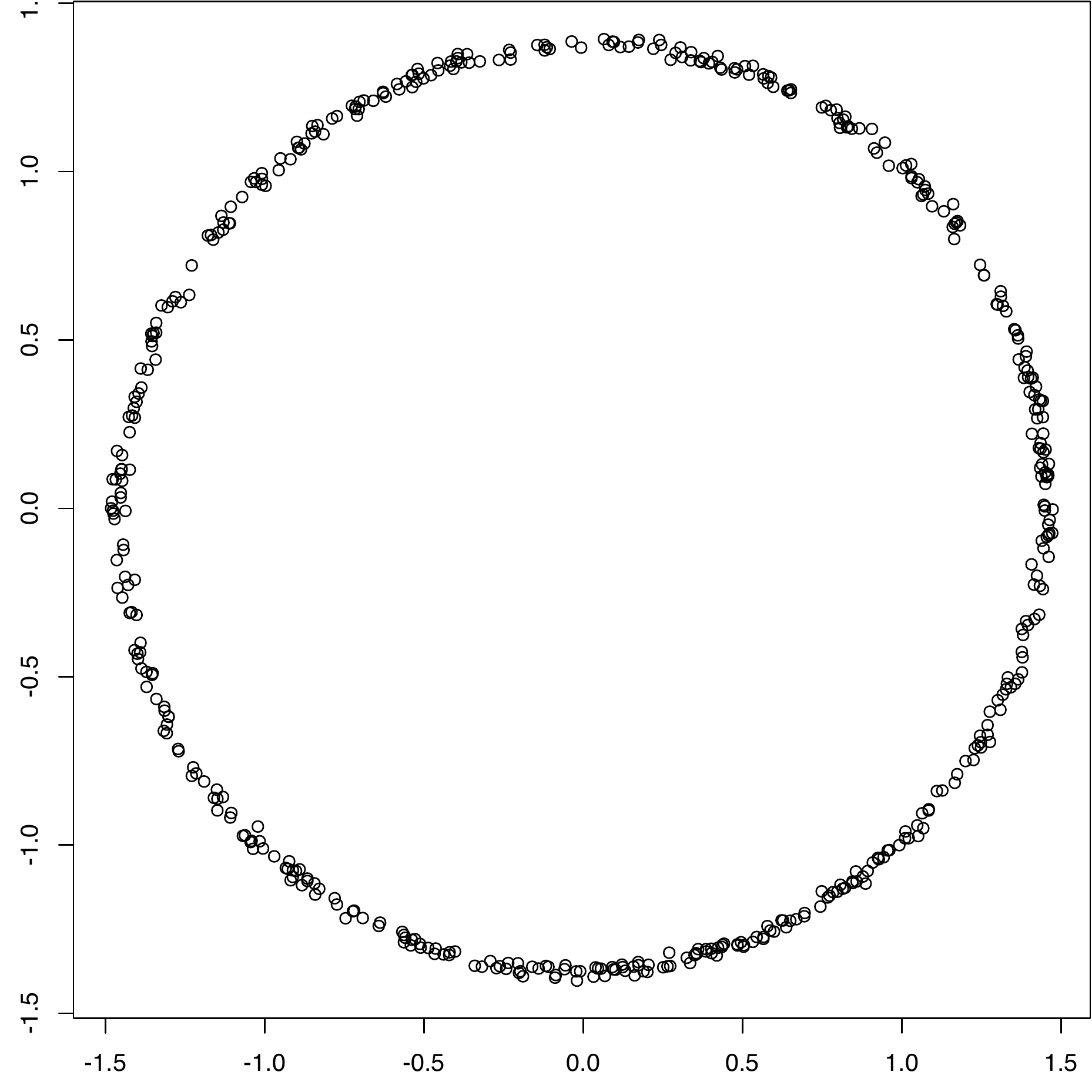}

\caption{PP for Example~2: Two-dimensional projections of the data $\x_i^{\rm ics}$, starting from components $(j,k)=(1,2)$ (upper left panel) and after $1$ (upper right panel), $2$ (lower left panel) and $4$ (lower right panel) iterations of local PP.}
\label{fig:Ex2B}
\end{figure}

Finally, we applied the procedure advocated in this manuscript. After running ICS with $\nu = 0$ and $\gamma = 1$, the initial values $\hat{H}(\bPi^\top \U_{jk}^\top \x_1^{\rm ics}, \ldots, \bPi^\top\U_{jk}^\top \x_n^{\rm ics})$ ranged from $2.7651$ to $2.8476$. The minimal value was obtained with $(j,k) = (1,2)$, and the corresponding scatter plot is shown in the upper left panel of Figure~\ref{fig:Ex2B}. Running local PP with this starting point revealed quickly the underlying structure. The total number of iterations was $13$, but already $4$ iterations gave away the uniform distribution on the circle, see the lower right panel of Figure~\ref{fig:Ex2B}.

This example illustrates nicely the benefit of using ICS as a means to find promising starting points for local PP.

\begin{figure}
\includegraphics[width=0.49\textwidth]{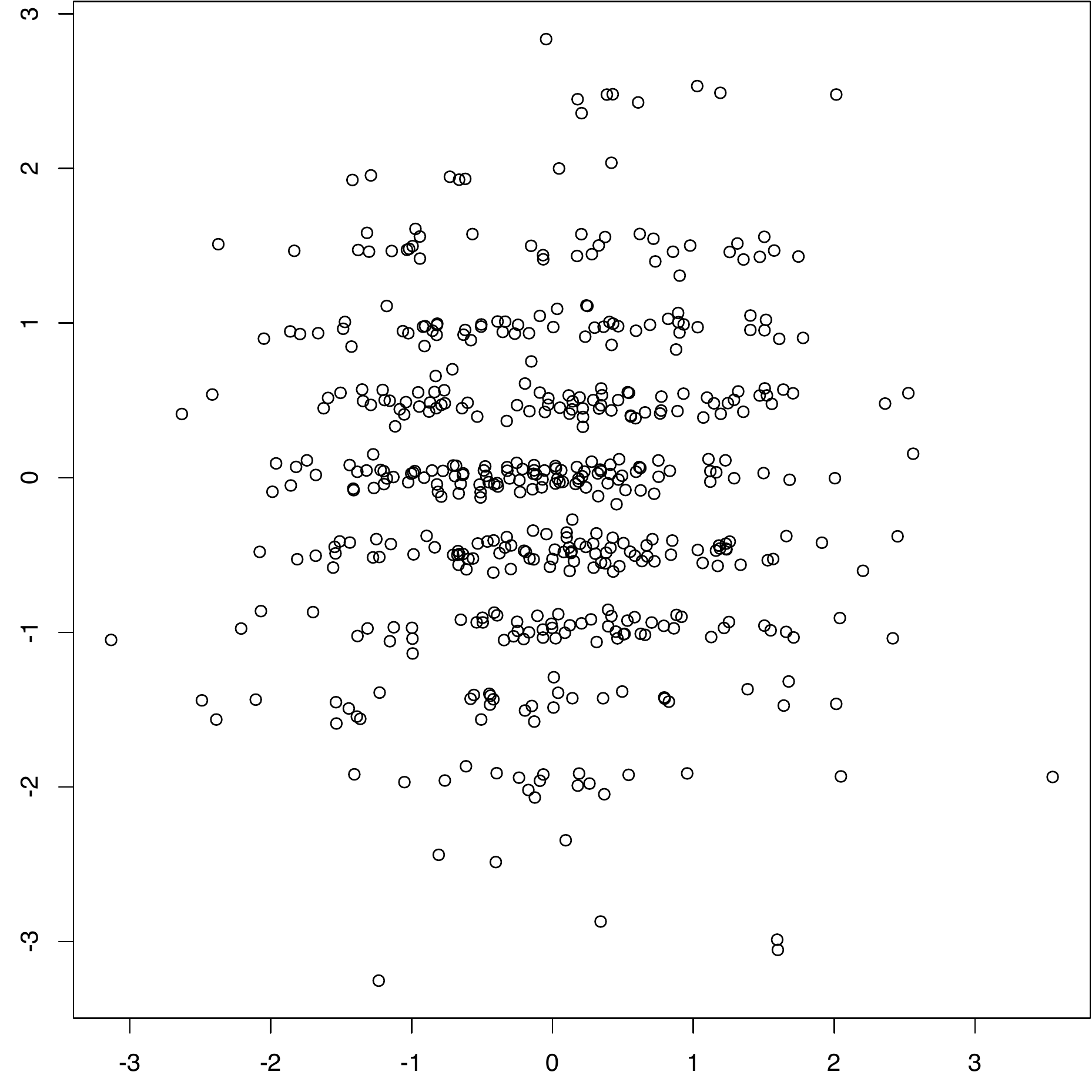}
\hfill
\includegraphics[width=0.49\textwidth]{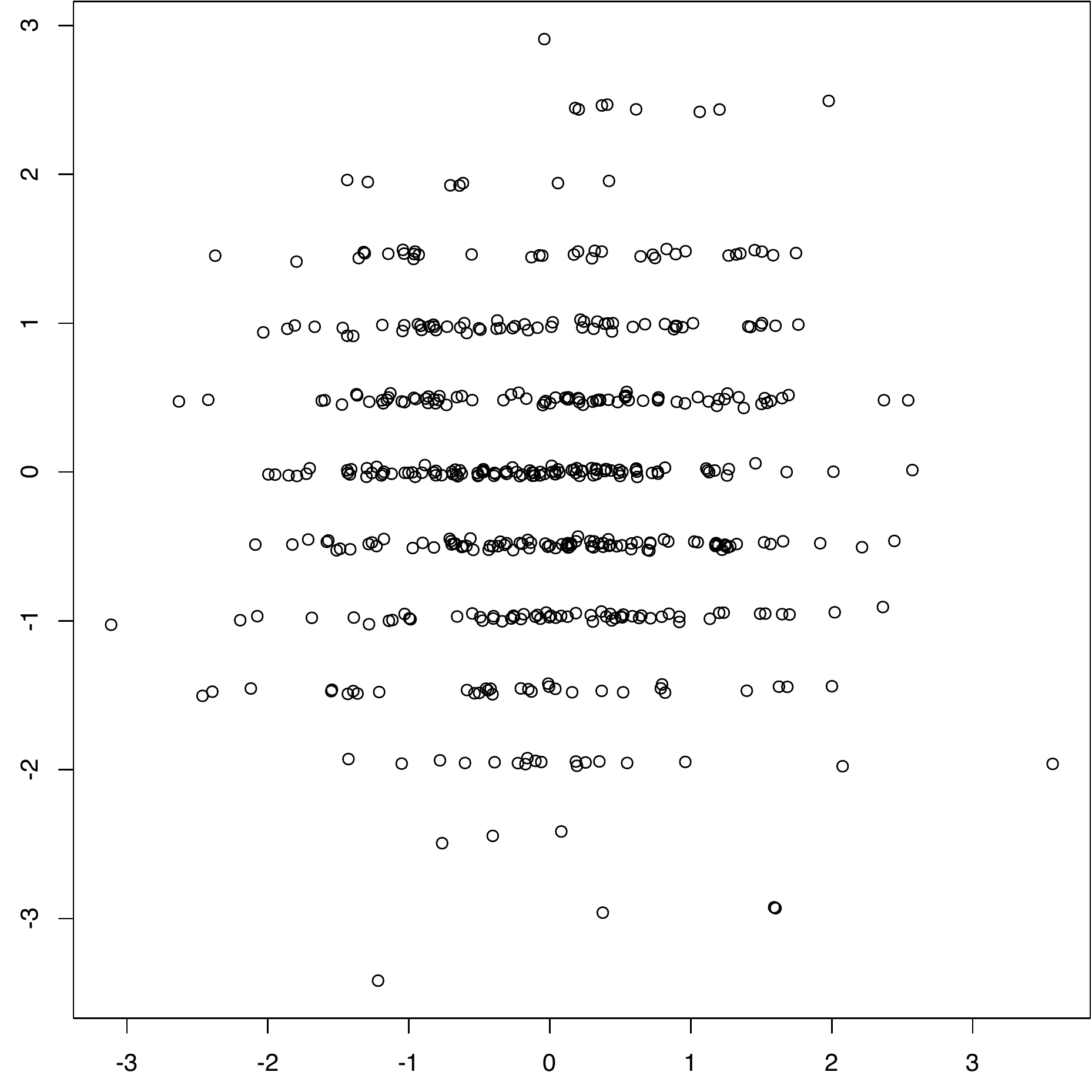}

\includegraphics[width=0.49\textwidth]{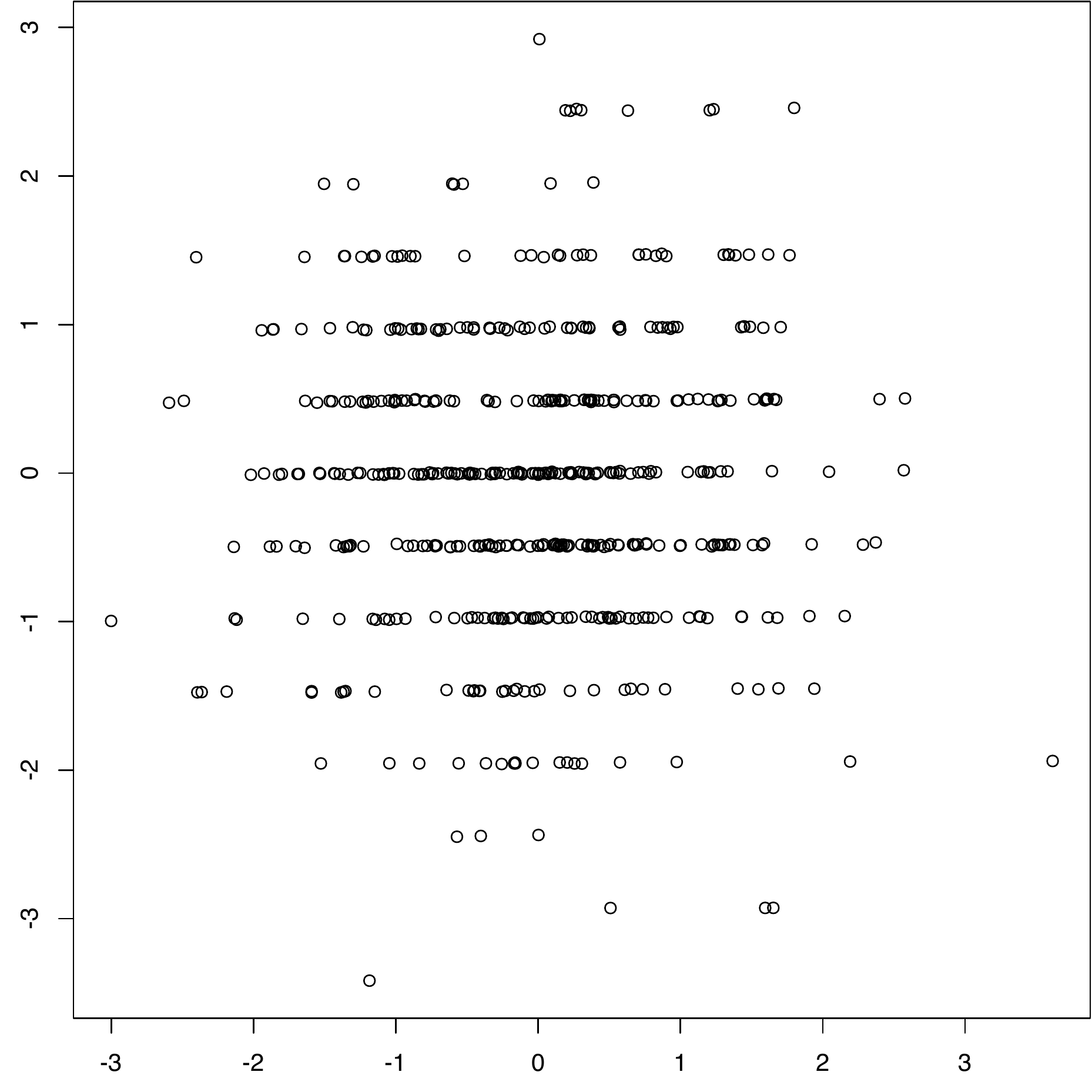}
\hfill
\includegraphics[width=0.49\textwidth]{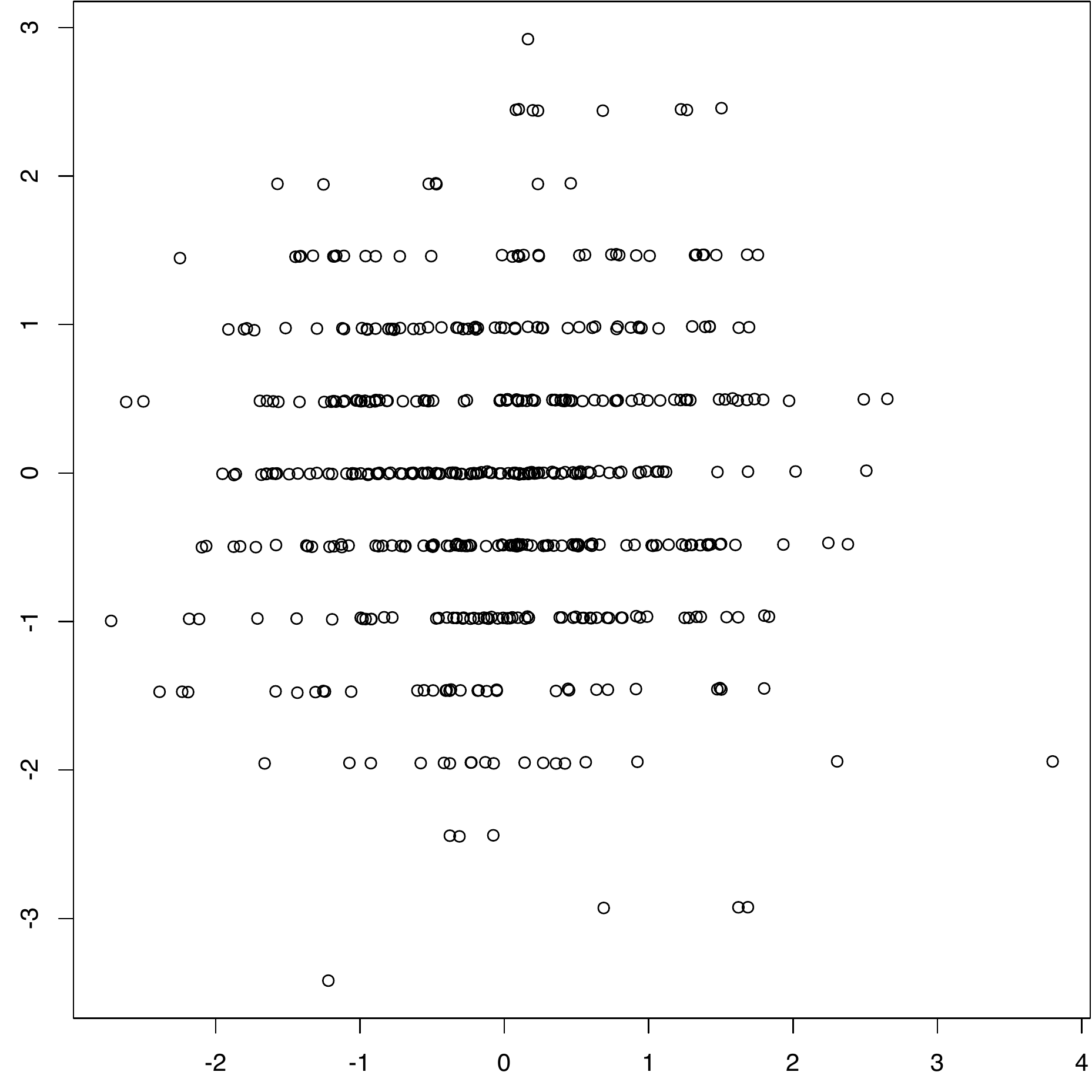}

\caption{PP for Example~3: Two-dimensional projections of the data $\x_i^{\rm ics}$, starting from components $(j,k)=(1,5)$ (upper left panel) and after $1$ (upper right panel), $4$ (lower left panel) and $85$ (lower right panel) iterations of local PP.}
\label{fig:Ex3}
\end{figure}

\paragraph{Example~3.}
Our final example concerns a structure which is suprisingly difficult do detect even in moderate dimension. Here the dimension is $p = 6$. Starting local PP starting from all $p(p-1)/2 = 15$ pairs of two components of the prewhitened data $\x_i^{\rm pre}$ did not reveal anything, neither for $h = 0.5$ nor for $h = 0.2$. Also our three-stage procedure with $\nu = 0$ and $\gamma = 1$ led nowhere. But with $\gamma = 4$, an interesting structure became visible for components $(j,k) = (1,5)$ of the observations $\x_i^{\rm ics}$, see the upper left panel of Figure~\ref{fig:Ex3}. The inital value of $\hat{H}$ was $2.6202$. After $85$ iterations of local PP, we ended up with the projection shown in the lower right panel, and the estimated entropy was $2.5797$.

\noindent
\textbf{Acknowledgements.} \
Part of this research was supported by Swiss National Science Foundation. The authors are grateful to two referees for constructive comments.

\end{document}